\documentclass[a4paper,twocolumn,11pt,accepted=2022-11-30]{quantumarticle}                            
\pdfoutput=1
\usepackage[utf8]{inputenc}
\usepackage[english]{babel}
\usepackage[T1]{fontenc}
\usepackage{amsmath}
\usepackage{hyperref}
\usepackage{lipsum}
\usepackage{eurosym}                                                                                
\usepackage{amsmath}
\usepackage{amsfonts}
\usepackage{amssymb}
\usepackage{graphics}
\usepackage{graphicx}
\usepackage{color}              
\usepackage{hyperref}  
\usepackage[numbers,sort&compress]{natbib}
\usepackage{braket}                                                                                 
\usepackage{accents}                                                                                
\usepackage{url}
\usepackage{textcase}
\usepackage{bm}
\usepackage[T1]{fontenc}
\usepackage[final]{microtype}
\usepackage{tikz}
\usepackage{mathtools}
\usepackage{soul}
\usetikzlibrary{arrows.meta}
\newcommand{\mi}{\mathrm{i}} 
\DeclareUnicodeCharacter{2212}{-}
\DeclareUnicodeCharacter{2009}{~}
\title{Variational Quantum Simulation of Valence-Bond Solids}
\author{Daniel Huerga}
\orcid{0000-0003-3210-2370}
\email{daniel.huerga@ubc.ca}
\affiliation{Stewart Blusson Quantum Matter Institute, University of British Columbia, Vancouver V6T 1Z4, BC, Canada}
\affiliation{Department of Physical Chemistry, University of the Basque Country UPV/EHU, Apartado 644, 48080 Bilbao, Spain}
\begin{document}
\begin{abstract}

We introduce a hybrid quantum-classical variational algorithm to simulate ground-state phase diagrams of frustrated quantum spin models in the thermodynamic limit. 
The method is based on a cluster-Gutzwiller ansatz where the wave function of the cluster is provided by a parameterized quantum circuit whose key ingredient is a two-qubit real XY gate allowing to efficiently generate valence-bonds on nearest-neighbor qubits. 
Additional tunable single-qubit Z- and two-qubit ZZ-rotation gates allow the description of magnetically ordered and paramagnetic phases while restricting the variational optimization to the U(1) subspace.
We benchmark the method against the $J_1$--$J_2$ Heisenberg model on the square lattice and uncover its phase diagram, which hosts long-range ordered N\'eel and columnar anti-ferromagnetic phases, as well as an intermediate valence-bond solid phase characterized by a periodic pattern of 2$\times$2 strongly-correlated plaquettes.
Our results show that the convergence of the algorithm is guided by the onset of long-range order, opening a promising route to synthetically realize frustrated quantum magnets and their quantum phase transition to paramagnetic valence-bond solids with currently developed superconducting circuit devices.
\end{abstract}
\maketitle

Hybrid quantum-classical variational algorithms, so-called variational quantum algorithms (VQA), are at the center of current research for their potentialities in providing \textit{useful} applications of currently developed noisy intermediate scale quantum (NISQ) devices \cite{Preskill_2018}.
They consist in a generic feedback loop where the NISQ device provides a quantum state via a parameterized quantum circuit (PQC) that is tuned by a classical computer so as to optimize a certain objective function encoding the problem of interest \cite{McClean_2016,Cerezo_2021_rev,Bharti_2021}.
The variational quantum eigensolver \cite{Peruzzo_2014} was one of the first VQA proposed to approximate the ground-state and energy of \textit{finite} strongly-correlated fermionic Hamiltonians as an alternative to the phase estimation algorithm \cite{Nielsen_Chuang_2010}, which provides the exact ground-state solution but requires coherence times on the quantum devices unreachable with current technology.

The initially expected unleashed potentialities of VQA towards providing quantum advantage on problems including machine learning, optimization, and the simulation of strongly-correlated electron systems ---a foundational motivation driving research in quantum computation \cite{Feynman_1982,Abrams_1997,Ortiz_2001,Somma_2002,Wecker_2015_Progress,Wecker_2015_Solving,Jiang_2018}--- have been narrowed down due to the identification of various limitations.
Specifically, the optimization landscape has been shown to be plagued with the so-called \textit{barren plateaux} \cite{McClean_2018}, large flat regions hindering optimization that may appear irrespective of the optimization routine used \cite{Arrasmith_2020} and induced by noise \cite{Wang_2021}.
Additionally, suboptimal minima have been argued to render the classical optimization problem NP-hard \cite{Bittel_2021}.
In spite of these limitations, \textit{local} objective functions, e.g. the energy of a local Hamiltonian, may still be efficiently optimized for shallow \textit{inexpressive} PQCs in certain regimes \cite{Cerezo_2021,Holmes_2022,Wang_2021}.

In particular, two-dimensional (2D) frustrated quantum magnets, characterized by the impossibility of finding a spin arrangement satisfying all local energy constraints simultaneously in any locally-rotated basis \cite{Lacroix_2011}, provide a natural testbed arena for VQAs and NISQ devices.
They pose a challenge to state-of-the-art classical numerical methods \cite{Hatano_1992,Troyer_2005,Marvian_2019} and at the same time host a plethora of phases and phenomena of both fundamental and applied interest.
In particular, quantum paramagnetic phases either breaking or preserving translational symmetry  ---so-called valence-bond solids (VBS) and quantum spin-liquids (QSL), respectively---, have important implications for layered materials \cite{Norman_2016,Zayed_2017,Zhou_2017} and quantum computation \cite{Verstraete_2004,Wei_2011,Miyake_2011,Kitaev_2003,Kitaev_2006}.
Recent developed hybrid approaches have mainly focused on the simulation of 1D lattices \cite{Schoen_2005,Kokail_2019,Foss-Feig_2020,Barratt_2021,Haghshenas_2022}, while approaches to 2D have been more scarce and limited to finite systems \cite{Liu_2019,Haghshenas_2022}.

Here, we introduce a cluster-Gutzwiller VQA to simulate ground-state phase diagrams of frustrated 2D quantum spin models \textit{in the thermodynamic limit}.
We build upon the grounds of hierarchical mean-field theory (HMFT) \cite{Batista_2004}, an algebraic framework based on the use of clusters for which a Gutzwiller ansatz represents the lowest order approximation. 
Furnished with a scaling analysis, it allows to uncover ground-state phase diagrams in the thermodynamic limit characterized by co-existence and competition of different long-range orders (LROs) and quantum paramagnetic phases \cite{Isaev_2009_JQ,Isaev_2009_SS,Isaev_2009_J1J2,Huerga_2013,Huerga_2014,Huerga_2016}.
Aiming at overcoming the scaling limitations of HMFT, here we present its \textit{quantum-assisted} approach, dubbed Q-HMFT, where the cluster wave function is generated via a PQC whose central element is a parameterized real two-qubit XY gate that efficiently generates valence-bonds on nearest-neighbor (NN) qubits.
Respecting the planar square connectivities of currently developed chips \cite{Arute_2019,Krinner_2022}, we provide a systematic PQC construction resulting in shallower depths than other commonly used PQC ansatze, which either admix symmetry sectors \cite{Bravo_Prieto_2020,Kandala_2017} or require higher chip connectivities \cite{Wecker_2015_Progress}.
Information about the thermodynamic limit is implemented at the objective function level (i.e. the energy) through the self-consistent mean-field embedding concomitant to the cluster-Gutzwiller ansatz.
Such mean-field embedding push the convergence of Q-HMFT within LRO phases, while paramagnetic phases are accessed by smoothly tuning the system through a quantum phase transition.
We benchmark numerically Q-HMFT on the paradigmatic antiferromagnetic $J_1$--$J_2$ model on the square lattice \cite{Chandra_1988,Dagotto_1989}, which hosts N\'eel and columnar anti-ferromagnet (CAF) LROs, as well as a long-dabeted intermediate paramagnetic phase \cite{Singh_1990,Read_1991,Capriotti_2000,Mambrini_2006,Darradi_2008,Isaev_2009_J1J2,Richter_2010,Jiang_2012,Yu_2012,Hu_2013,Wang_2013,Gong_2014,Morita_2015,Wang_2016,Wang_2018}.
Consistently with previous classical results from HMFT \cite{Isaev_2009_J1J2} and other state-of-the-art algorithms \cite{Capriotti_2000,Mambrini_2006,Yu_2012,Gong_2014}, we find that the N\'eel melts onto a VBS characterized by a 2$\times$2 \textit{plaquette} ordering.
Convergence results from up to a 4$\times$4-qubit cluster suggest the potential scalability of Q-HMFT, making it a promising route to quantum simulate frustrated magnets and valence-bond solid phases with currently developed superconducting qubit devices beyond classical capabilities.

\section{Quantum-assisted hierarchical mean-field} 
Let us consider a generic translational invariant 2-body local Hamiltonian in the infinite lattice, $H$=$\sum_{i,j} h_{i,j}$, where $h_{i,j}$=$J_{i,j}\mathbf{S}_i\mathbf{S}_j$ and $\mathbf{S}_i$=$(S^x_i,S^y_i,S^z_i)$ refers to the S=1/2 SU(2) spin operators at site $i$, and $J_{i,j}$ to the two-spin interaction strength.
We tile the lattice with $N$-site equivalent clusters preserving as much as possible the original symmetries of the lattice.
We classify the Hamiltonian terms on those acting within clusters, $h^{\square}_R=\sum_{(i,j)\in R}h_{i,j}$, and those acting on two different clusters, $h^{I}_{R,R'}= \sum_{i\in R,j\in R'} h_{i,j}$, where $R$ labels the position of the cluster in the tiled lattice, or superlattice.
We define a uniform cluster-Gutzwiller ansatz as the product state of $N$-site clusters in the infinite lattice, 
\begin{equation}
\ket{\Psi({\bm{\theta}})} = \bigotimes_{R}\ket{\psi({\bm{\theta}})}, \label{eq:Psi}
\end{equation}
where we have restricted all clusters $R$ to be in a same normalized state, $\ket{\psi({\bm{\theta}})}$, parameterized by $\bm{\theta}$=$\lbrace \theta_k\rbrace$.
Considering the wave function \eqref{eq:Psi}, the energy per spin reduces to the contribution of a single cluster and its mean-field embedding \cite{Isaev_2009_J1J2,Huerga_2013,Huerga_2014},
\begin{equation}
E({\bm{\theta}})= \frac{1}{N}\left(\langle h^{\square} \rangle_{\bm{\theta}}
+ \frac{1}{2}\sum_{i\in \square,j\in \square'} J_{i,j}\langle \mathbf{S}_i \rangle_{\bm{\theta}} \langle \mathbf{S}_j \rangle_{\bm{\theta}}\right)
, \label{eq:energy}
\end{equation}
where $\langle \cdot \rangle_{\bm{\theta}}$ refers to the expectation value with the parameterized cluster wave function, $\ket{\psi(\bm{\theta})}$.
The first summand in \eqref{eq:energy} accounts for the intra-cluster contributions, while the second runs over all two-spin inter-cluster interactions with the neighboring clusters, $\square'$ (the 1/2 preventing double counting) and provides information about the thermodynamic limit by allowing the explicit breakdown of symmetries and concomitant onset of LROs.
Generalization to $n$-body interactions appearing in e.g. ring-exchange models is straightforward \cite{Isaev_2009_JQ,Huerga_2014,Huerga_2017}. 
The optimal energy $E$=$E({\bm{\theta^\star}})$ with parameters $\bm{\theta}^\star$ is obtained upon minimization of the variational energy \eqref{eq:energy}.
Generically, the exact limit is recovered at $N$$\rightarrow$$\infty$. 
Thus, by increasing the cluster size, we assess the stability of the phase diagram. 
However, exact diagonalization of the clusters is typically limited to $N$$\lesssim$$30$ with classical computing means, which conforms the major bottleneck of the method \cite{Isaev_2009_J1J2,Isaev_2009_SS,Huerga_2013,Huerga_2016,Huerga_2017}. 
%
%
\begin{figure}[t]
\includegraphics[scale=0.74]{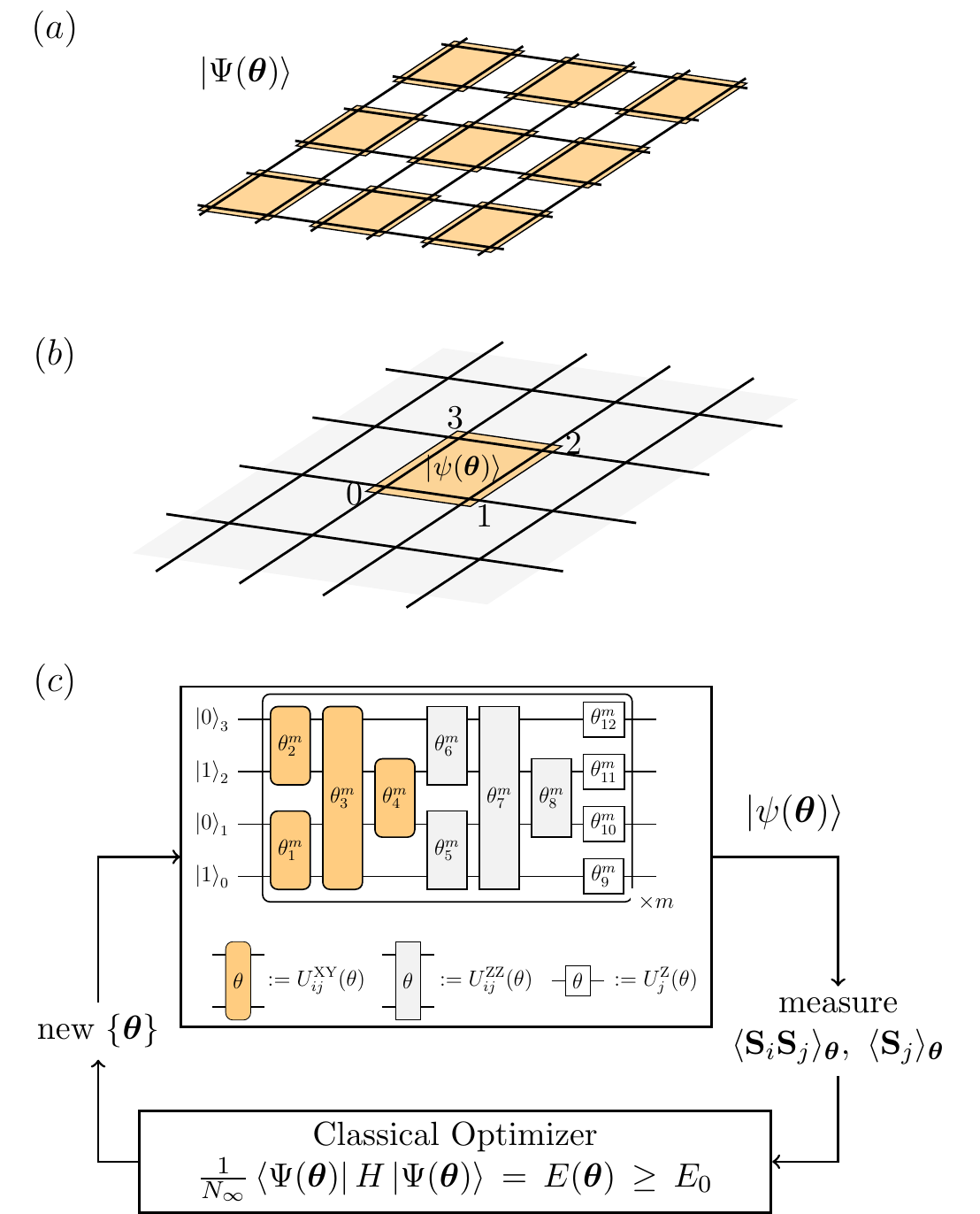}
\caption{
\label{fig:method}
($a$) Schematic representation of the cluster-Gutzwiller ansatz \eqref{eq:Psi} with 2$\times$2 clusters. The variational equations are equivalent to solving a single cluster in a self-consistent mean-field embedding, represented in gray shade in ($b$). ($c$) Q-HMFT loop: The wave function of the cluster is provided by a PQC whose parameters $\lbrace \bm{\theta}\rbrace$ are tuned so as to optimize the energy density in the thermodynamic limit \eqref{eq:energy}, which is an upper bound to the exact, $E_0$.
}
\end{figure}

In order to overcome such limitation, here we consider a PQC generating the cluster wave function,
\begin{equation}
\ket{\psi({\bm{\theta}})}= \mathcal{U}({\bm{\theta}})\ket{\psi_0},\label{eq:psi_cluster}
\end{equation}
where $\ket{\psi_0}$ is an initial easy-to-prepare state, here onwards fixed to $\ket{\psi_0}=\ket{0101\ldots 01}$ for reasons that will be clear below, with $\ket{0}\doteq\ket{\uparrow},~\ket{1}\doteq\ket{\downarrow}$, and $\mathcal{U}({\bm{\theta}})$ a digitalized unitary transformation, $\mathcal{U}({\bm{\theta}})=\prod_{k} U_k(\theta_k)$ where $U_k(\theta_k)=e^{-\mi \theta_k V_k}$ with $V_k$ an hermitian operator acting on either one or two qubits (see Fig.\ref{fig:method}).

For SU(2) Hamiltonians on a bipartite lattice, we may restrict the variational search to the U(1) subspace of null total magnetization, $\sum_{j\in\square}\langle S^z_j\rangle$=0.
Within this subspace, the self-consistent mean-field embedding is restricted to the $z$-magnetization, i.e. $\langle S^x_j\rangle$=$\langle S^y_j\rangle$=0.
This restriction allows to reduce the variational search and the PQC depth while still allowing for the description of symmetry breaking (long-range ordered magnetic) solutions, as well as SU(2) preserving (paramagnetic) ones.
Within the two-qubit $S^z$=0 subspace, spanned by $\ket{\tilde{0}}$$\doteq$$\ket{10}$ and $\ket{\tilde{1}}$$\doteq$$\ket{01}$, any transformation can be generated by the XY-Heisenberg, $\widetilde{S}_{ij}^x$=$(S^x_i S^x_j + S^y_i S^y_j)$, and the $z$-projection of the Dzialoshinskii-Moriya (DM), $\widetilde{S}_{ij}^y$=$(S^x_i S^y_j - S^y_i S^x_j)$, interactions. 
Together with $\widetilde{S}_{ij}^z$=$\frac{1}{2}(S^z_j - S^z_i)$, they form a SU(2) algebra, $[\widetilde{S}_{ij}^\alpha,\widetilde{S}_{ij}^\beta]$=${\rm i}\varepsilon^{\alpha\beta\gamma}\widetilde{S}^\gamma_{ij}$.
In particular, a rotation generated by the DM interaction, $V_{ij}=\widetilde{S}^y_{ij}$, leads to a purely real XY-gate,
\begin{eqnarray}
U^{\text{XY}}_{ij}(\theta)
&=&
\cos(\theta/2)(
\ket{01}\bra{01}+\ket{10}\bra{10})
\notag\\
    &+&
\sin(\theta/2)(
\ket{01}\bra{10}-\ket{10}\bra{01})
,\label{XY_gate}
\end{eqnarray}
which efficiently transforms an uncorrelated N\'eel into a singlet, i.e. a \textit{valence-bond}, $U^{\text{XY}}_{ij}\left(\frac{\pi}{2}\right)\ket{01}_{ij}=\frac{1}{\sqrt{2}}(\ket{01}_{ij}-\ket{10}_{ij}).$ 
The {\sf REAL}-XY gate \eqref{XY_gate} can be considered as a generalization of a Givens rotations \cite{Golub_1989} to Hilbert spaces, and its use has been proposed for the digital preparation of fermionic Slater determinants \cite{Wecker_2015_Solving,Jiang_2018}. 
Recently, a controlled version of it has been argued to be universal \cite{Arrazola_2022}.
From the experimental standpoint, recent implementations have shown fidelities $\sim 97\%$ \cite{Abrams_2020}.

Considering the planar square connectivities of currently developed superconducting quantum chips \cite{Arute_2019,Krinner_2022}, we use $L$$\times$$ L$ qubit clusters with even $L$.
We heuristically construct the PQC \eqref{eq:psi_cluster} as it follows.
First, we apply a set of {\sf REAL}-XY gates \eqref{XY_gate} to all pairs of NN qubits in an order that (i) packs as much as possible two-qubit gates per layer and (ii) favors the C$_4$ symmetry of the square lattice.
Second, in order to tune the sign-structure of the resulting wave function and increase the amount of correlations, we add two-qubit parameterized ZZ-rotations, $ U^{\text{ZZ}}_{ij}(\theta)$=$e^{-\mi \theta Z_i Z_j}$ where $Z_j$=$2S^z_j$, in the same order described before. 
Finally, we add a layer of single-qubit parameterized Z-rotation gates, $U^{\text{Z}}_j(\theta)$=$e^{-\mi \theta Z_j}$.
Both ZZ- and Z-rotations are realizable experimentally \cite{NLacroix_2020}.
Such a XY-ZZ-Z macro-layer, which minimizes the circuit depth within the considered gateset, is recursively applied a few times $m$ to improve accuracy.
Specifically, 
\begin{equation}
\mathcal{U}= \prod_m \left(\mathcal{U}^\text{Z}\prod_{g=\text{XY,ZZ}}\left[T_y\mathcal{U}_y^{g}\right]\left[T_x\mathcal{U}_x^{g}\right]\mathcal{U}_y^{g}\mathcal{U}_x^{g}\right),
\label{eq:layers}
\end{equation}
where we have defined
$
\mathcal{U}_\alpha^{g}$=$\prod_{k} U^{g}_{(\mathbf{r}_k,\mathbf{r}_k+\hat{e}_\alpha)}(\theta_k^m)
$ 
as the product of $g$-type two-qubit gates applied to NN dimers in a columnar pattern along $\alpha$=$\lbrace x,y\rbrace$, where $\mathbf{r}_k$=$(x_k,y_k)$ labels the position of the qubit considering the bottom-left corner as the origin, and $\hat{e}_\alpha$ is the unit vector. 
That is, $0\le x_k \le L-2$ (for even $x_k$) and $0 \le y_k \le L-1$ for $\mathcal{U}_x^g$, and vice versa for $\mathcal{U}_y^g$.
The operator $T_\alpha \mathcal{U}_\alpha^{g}$ refers to the translation of $\mathcal{U}^g_{\alpha}$ by $\hat{e}_\alpha$, and $\mathcal{U}^\text{Z}$=$\prod_{j} U^{\text{Z}}_{\mathbf{r}_j}(\theta_j^m)$ is the last single-qubit layer.
For $L$=2, no translation operators are needed to uncover all NNs, thus $d$$=$$5m$ (see Fig.\ref{fig:method}).
For $L$$\ge$4, the PQC \eqref{eq:layers} has $n$=$(5L^2-4L)m$ variational parameters and a depth $d$=$9m$ (see Supplementary Material). 
As shown in the following, a small $m$=2 is sufficient to exactly reproduce classical HMFT with 2$\times$2 clusters, and provides with a good quantitative approximation to HMFT-4$\times$4 results.

\section{Numerical results}
We benchmark Q-HMFT on the antiferromagnetic $J_1$--$J_2$ Heisenberg model on the square lattice \cite{Chandra_1988,Dagotto_1989},
\begin{equation}
H= J_1\sum_{\langle i,j\rangle} \mathbf{S}_i\mathbf{S}_j + 
J_2\sum_{\langle\langle i,j\rangle\rangle} \mathbf{S}_i\mathbf{S}_j,
\label{eq:J1-J2}
\end{equation}
where $\langle i,j \rangle$ and $\langle\langle i,j\rangle\rangle$ account for NN and next-to-NN pairs of sites, respectively.
From here onwards, we express all quantities in units of $J_1$.
Upon tuning $J_2>0$, model \eqref{eq:J1-J2} hosts N\'eel and CAF ordered phases, and an intermediate quantum paramagnetic phase in between that is a matter of debate \cite{Singh_1990,Read_1991,Capriotti_2000,Mambrini_2006,Darradi_2008,Isaev_2009_J1J2,Richter_2010,Jiang_2012,Yu_2012,Hu_2013,Wang_2013,Gong_2014,Morita_2015,Wang_2016,Wang_2018}. 
Proposals have fluctuated through the time, including different QSLs \cite{Jiang_2012,Wang_2013,Hu_2013}, and VBSs with a pattern of \textit{dimers} \cite{Singh_1990,Read_1991,Morita_2015} or 2$\times$2 \textit{plaquettes} \cite{Capriotti_2000,Yu_2012,Gong_2014} breaking or preserving C$_4$ symmetry, respectively, the last scenario being described by HMFT \cite{Isaev_2009_J1J2}.
Interestingly, upon introducing third-NN Heisenberg interaction, the plaquette-VBS order becomes stronger \cite{Mambrini_2006} and has been argued to be of a higher-order symmetry-protected topological type \cite{Gonzalez-Cuadra_2022}.

We perform ideal noiseless classical simulations of Q-HMFT on the $J_1$--$J_2$ model \eqref{eq:J1-J2} with 2$\times$2 and 4$\times$4 clusters, and $m$=2 and 4 macro-layers. 
We compute energy, magnetization and dimer observables, as well as provide details on its convergence, and compare with the classical HMFT results with the same clusters, which provide with a satisfactory extrapolation to the $N$$\rightarrow$$\infty$ limit, $E_\infty(J_2=0)$$\simeq$$-0.64$ \cite{Isaev_2009_J1J2}, when compared to that obtained with state-of-the-art computational techniques, $E_\infty(J_2=0)$$\simeq$$-0.67$ \cite{Trivedi_1989,Richter_2010}.
As a classical optimizer, we use the gradient-based L-BFGS-B algorithm \cite{Byrd_1995,Zhu_1997}, where the gradient of the energy is computed via two-point first order approximation,
$\partial_{\theta_k} E= (E_{\mathbf{\theta}_k+\delta}-E_{\mathbf{\theta}_k})/\vert\delta\vert$, for $\delta=10^{-10}$ along $k$ direction in variational space.
We uncover the phase diagram by first obtaining an optimal solution out of various random PQC initializations at each extreme (i.e. $J_2$=0 and $J_2$=1), and then smoothly tuning the Hamiltonian parameters on both directions. 
The optimal parameters of one value of $J_2$ are used as initial parameters in the optimization of the immediate following point in the phase diagram.
We compute the energy \eqref{eq:energy} and its derivatives to detect quantum phase transitions, as well as magnetic and dimer orders to characterize the phases. 

%
%
\begin{figure}[t]
\includegraphics[width=0.5\textwidth]{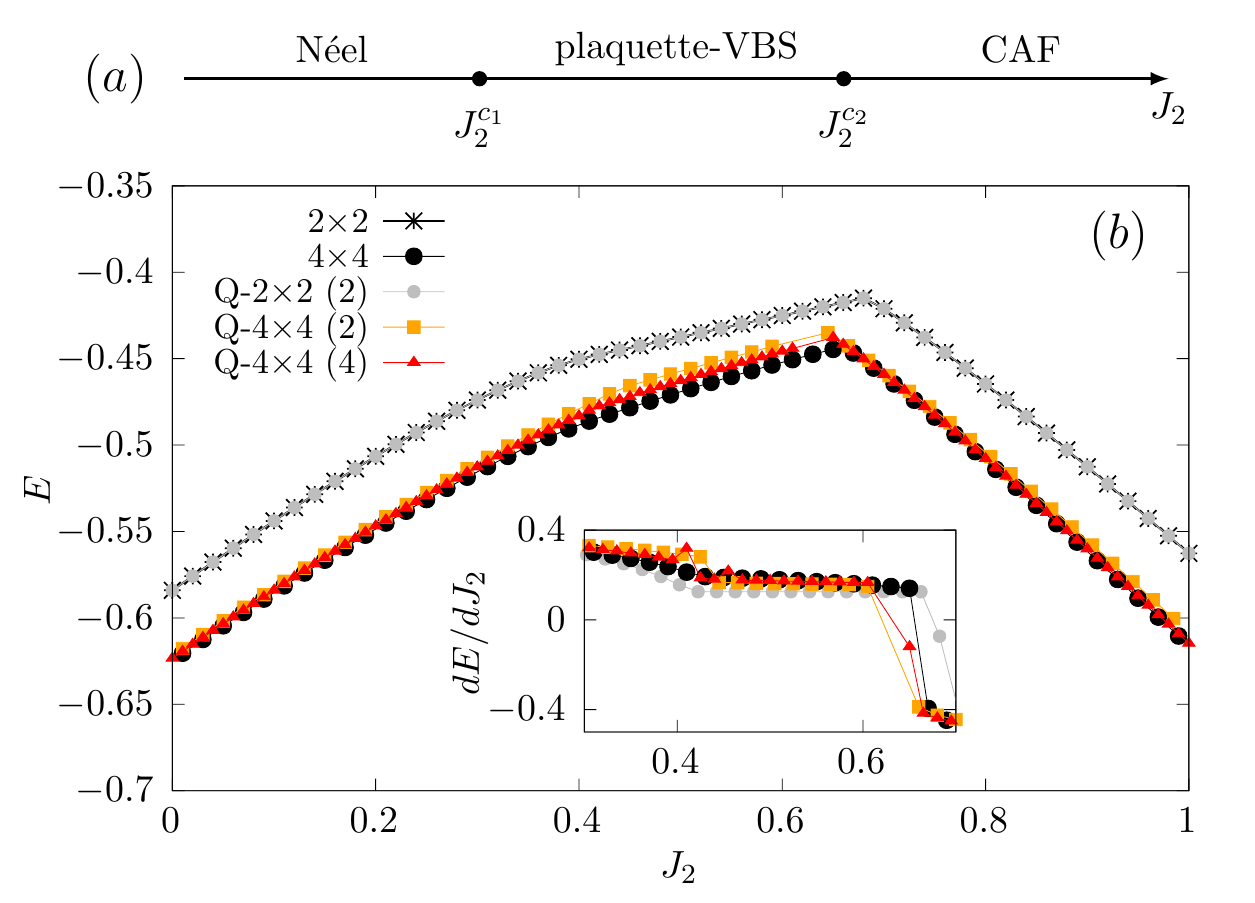}
\caption{\label{fig:energy}
$(a)$ Schematic phase diagram of the $J_1$--$J_2$ model \eqref{eq:J1-J2} as computed with hybrid and classical versions of HMFT-$L$$\times$$L$.
$(b)$ Energy per spin \eqref{eq:energy} and first derivative (inset) computed with Q-HMFT-$L$$\times$$L$ and $m$ macro-layers, labeled Q-$L$$\times$$L$ ($m$), together with HMFT results (labeled $L$$\times$$L$) for $L$=2, 4, and $m$=2, 4.
Q-HMFT-2$\times$2 with $m$=2 reproduces exactly the classical HMFT-2$\times$2 energy and its derivative (not shown for clarity purposes).}
\end{figure}
In Fig. \ref{fig:energy}, we show the optimal energy per spin and first derivative with respect to $J_2$ as computed with Q-HMFT with $m$=2 and $m$=4 macro-layers, together with HMFT results.
For 2$\times$2 clusters, Q-HMFT with $m$=2 and fixing equal parameters within the XY and ZZ layers, respectively (i.e. with 12 variational parameters), reproduces exactly classical HMFT results and describes a N\'eel phase and its melting through second order quantum phase transition onto a plaquette-VBS at $J_2^{c_1}$$\simeq$0.42, and a posterior first order phase transition at $J_2^{c_2}$$\simeq$0.68.
For 4$\times$4, Q-HMFT provides an excellent approximation to HMFT, increasing its accuracy as we increase the number of macro-layers.
Specifically, it shows a weak first order transition at $J_2^{c_1}$$\simeq$0.44 that tends towards second order upon increasing $m$, while the transition point moves towards the one obtained by HMFT, $J_2^{c_1}$$\simeq$0.42.
At $J_2^{c_2}$$\simeq$0.64, it shows a clear first order transition, which moves to the HMFT result, $J_2^{c_2}$$\simeq$0.66, upon increasing $m$.

To characterize the phases, first we inspect the magnetization,
\begin{equation}
M^z(\mathbf{k})= \frac{1}{N}\sum_{j}e^{-\mi \mathbf{r}_j\mathbf{k}} \langle S^z_j\rangle, \label{eq:magnetization}
\end{equation}
where $\mathbf{r}_i=(r_i^x,r_i^y)$ refers to the position of site $i$ in the infinite lattice and $\mathbf{k}$ to a vector in the first Brillouin zone.
N\'eel and CAF LROs are signaled by finite $M^z(\pi,\pi)$ and $M^z(0,\pi)$ (equivalently, $M^z(\pi,0)$), respectively.
%
%
\begin{figure}[t]
\includegraphics[width=0.45\textwidth]{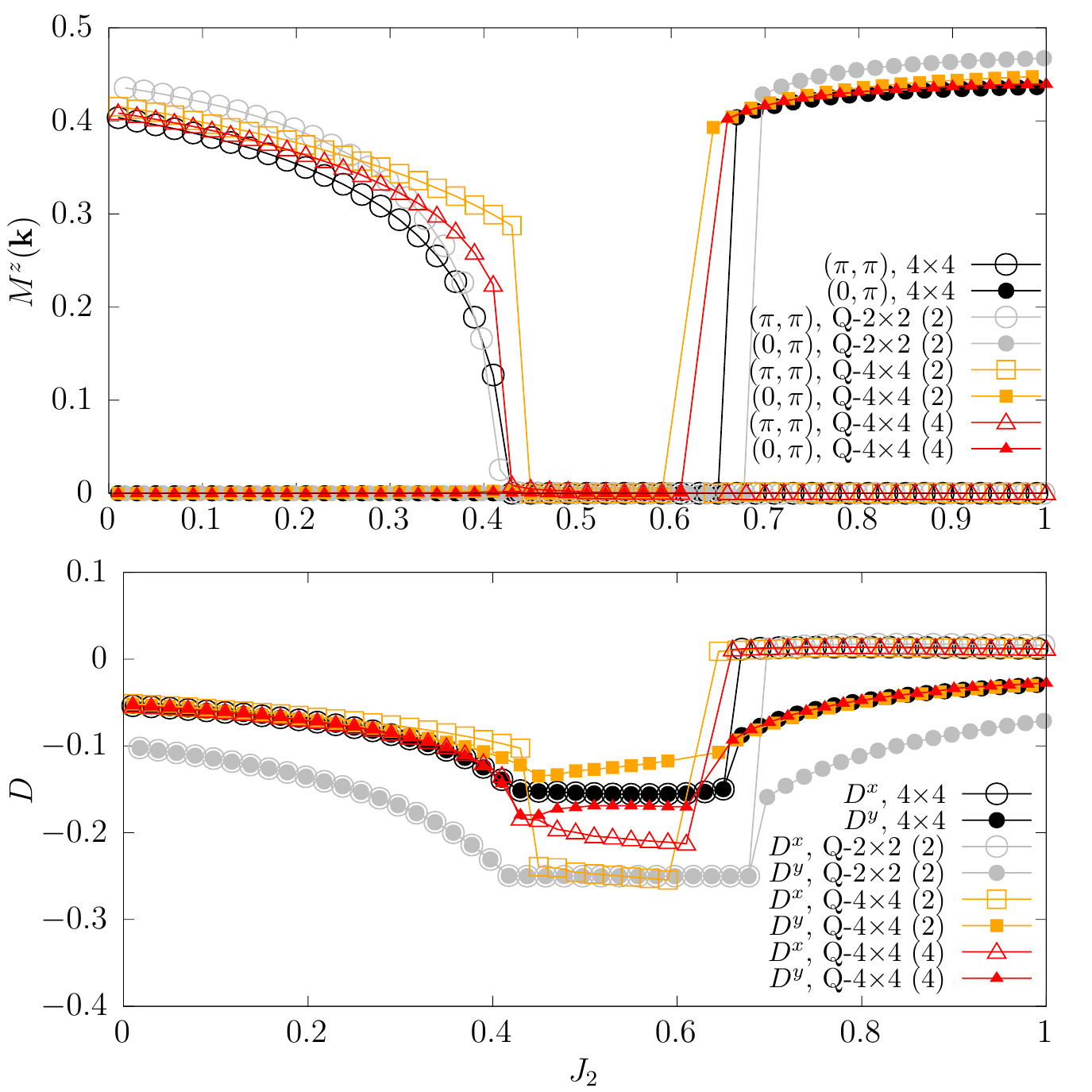}
\caption{\label{fig:magnetization_dimerization} Top: N\'eel (empty points) and CAF (filled points) LRO parameters computed with Q-HMFT-$L$$\times$$L$ and $m$ (labeled Q-$L$$\times$$L$ ($m$)), and HMFT results (labeled $L$$\times$$L$). Bottom: Dimer observable along $x$ (empty points) and $y$ (filled points) with Q- and HMFT.}
\end{figure}
In Fig. \ref{fig:magnetization_dimerization} (top) we show Q-HMFT results on the magnetization together with HMFT-4$\times$4 for comparison.
For 2$\times$2 clusters, Q-HMFT ($m$=2) reproduces exactly HMFT results (not shown for clarity purposes) showing the continuous vanishing of the N\'eel order parameter, an intermediate paramagnetic region, and a following discontinuous onset of CAF order, consistent with the description of second and first order phase transitions, respectively.
For 4$\times$4 clusters, Q-HMFT tends towards HMFT results upon increasing $m$, essentially providing the same phase diagram as with (Q-)HMFT-2$\times$2.
In particular, the vanishing of the N\'eel order parameter is discontinuous with $m$=2 (consistent with the first order transition seen in the energy), but such discontinuity decreases upon increasing $m$=4, suggesting a continuous vanishing for $m$$>$4 as found with HMFT.

In order to characterize the possible onset of VBS order in the intermediate paramagnetic phase, we define the dimer observable along $\alpha=x,y$ allowing to distinguish between dimer or plaquette order \cite{Capriotti_2000},
\begin{equation}
D^{\alpha}= \frac{1}{L}\sum_{\langle i,j\rangle_\alpha} (-1)^{r^\alpha_i}\langle \mathbf{S}_i \mathbf{S}_j\rangle,\label{eq:dimer}
\end{equation}
where $\langle i,j\rangle_\alpha$ refers to nearest-neighbor bonds along $\alpha$.
Specifically, $D^x$$\neq$$ D^{y}$ signals the onset of dimer-VBS along one direction, while $D^x$=$D^y$ provides a signature of a C$_4$ preserving plaquette-VBS.
In Fig. \ref{fig:magnetization_dimerization} (bottom) we show dimerization for Q-HMFT and its classical counterpart.
For 2$\times$2, Q-HMFT reproduces exactly HMFT results (not shown for clarity purposes) showing a non-zero $D^x$=$D^y$ within the N\'eel phase that increases in absolute value and reaches a plateaux indicating the plaquette-VBS at $J_2^{c_1}$.
Upon further tuning $J_2$, both dimer observables show a discontinuity at $J_2^{c_2}$ consistent with the first order transition found upon inspecting the energy. 
For 4$\times$4 clusters, HMFT essentially provides the same picture, but with a reduced value of the dimerization throughout the phase diagram.
Q-HMFT approximates quantitatively well HMFT results within the LRO phases, but shows a slight breakdown of the C$_4$ symmetry, i.e. $D_y$$<$$D_x$, within the intermediate plaquette-VBS.
Nevertheless, such difference is reduced upon increasing $m$, tending towards the HMFT-4$\times$4 results.

%
%
\begin{figure}[t]
\includegraphics[width=0.49\textwidth]{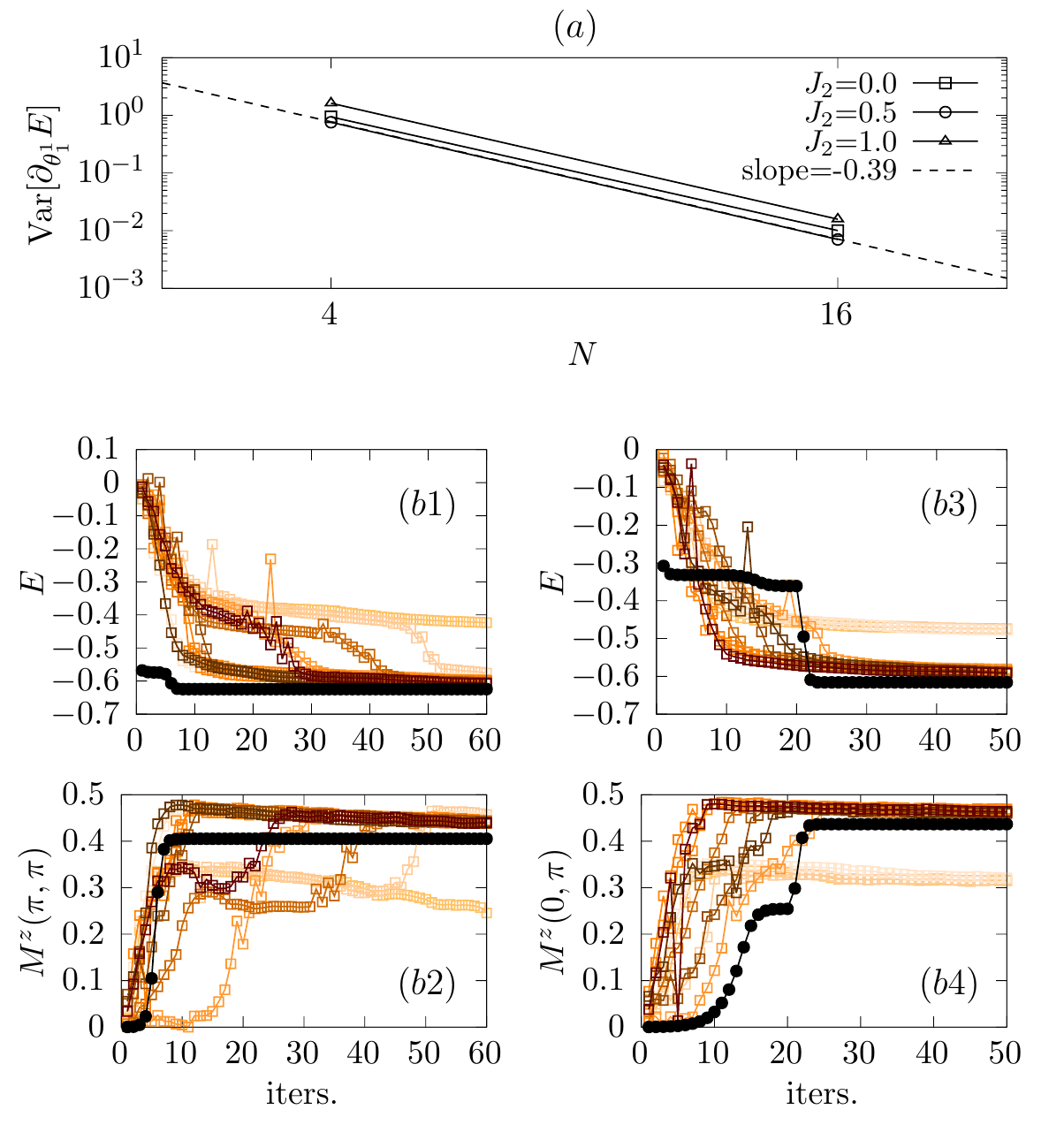}
\caption{($a$) Variance of the first component of the PQC for Q-HMFT, $m$=2. ($b$) Convergence of the energy and magnetization for Q-HMFT-4$\times$4, $m$=2, at $J_2$=0 ($b1$)--($b2$) and $J_2$=1 ($b3$)--($b4$) for 10 random initializations of the PQC (orange scale, empty squares) together with HMFT-4$\times$4 convergence results from a random initialization (black dots).}
\label{fig:convergence}
\end{figure}
Last, we analyze the scalability of Q-HMFT by inspecting the variance with respect to the first variational parameter, Var[$\partial_{\theta^1_1}E_{\bm{\theta}}$], estimated over 100 random initializations of the PQC, and the convergence of Q-HMFT-4$\times$4 with $m$=2 macro-layers.
As shown in Fig.\ref{fig:convergence}, the variance is strongly suppressed upon increasing the cluster size from 2$\times$2 to 4$\times$4 well within all three phases (i.e. at $J_2$=0, 0.5, and 1), $\text{Var}[\partial_{\theta^1_1}E](N=4)$$\sim$$O(1)$ and $\text{Var}[\partial_{\theta^1_1}E](N=16)$$\sim$$O(10^{-2})$.
Nevertheless, within the LRO phases, i.e. at $J_2$=0 (N\'eel) and $J_2$=1 (CAF), the convergence is pushed by the nonzero mean-fields concomitant to the onset of LRO.
The system may explore suboptimal minima, but in most of the cases some extra tens of iterations permit the system to escape from them and approximate the classical HMFT result.
On the contrary, in the quantum paramagnetic phase ($J_2$=0.5), which is characterized by a null mean-field embedding both in (Q-)HMFT for the $J_1$--$J_2$ model, the system explores a manifold of states characterized by different dimer orderings and magnetizations without arriving to converge (see Supplementary Material).

\section{Conclusion and outlook}
We have presented an hybrid algorithm to quantum simulate 2D quantum frustrated magnets in infinite lattices.
Based on the cluster-Gutzwiller ansatz, we present a quantum-assisted approach to HMFT \cite{Isaev_2009_J1J2,Huerga_2013,Huerga_2014,Huerga_2016}, dubbed Q-HMFT, where the wave function of a finite cluster is provided by a U(1) symmetry preserving PQC that respects the native NN square connectivities of currently developed superconducting quantum circuits, while the thermodynamic limit is accounted for by a self-consistent mean-field embedding introduced at the objective function level, i.e. the energy.
The main ingredient of the PQC is a parameterized {\sf REAL}-XY gate performing generalized Givens rotations in the two-qubit odd-parity subspace and efficiently generating valence-bonds.
Addition of parameterized ZZ- and Z-rotations furnish a fixed-depth macro-layer for general even $L$$\times$$L$-qubit clusters that is recursively applied a few $m$ times to increase accuracy.

We provide noiseless benchmark numerical results on the paradigmatic frustrated $J_1$--$J_2$ Heisenberg model with $L$=2 and 4 clusters, and $m$=2 and 4 macro-layers, and show that $m$=2 is sufficient to approach quantitatively well previous classical HMFT results \cite{Isaev_2009_J1J2}. 
In particular, $m$=2 \textit{exactly} reproduces HMFT-2$\times$2, describing a phase diagram hosting N\'eel and columnar antiferromagnetic (CAF) phases, as well as an intermediate quantum paramagnetic \textit{plaquette}-VBS phase, in accordance with results from state-of-the-art classical numerical techniques \cite{Capriotti_2000,Mambrini_2006,Gong_2014}.
With 4$\times$4 clusters, Q-HMFT provides an excellent approximation to HMFT within LRO phases, while the plaquette-VBS phase shows a slight dimerization ---and thus breakdown of C$_4$--- that is decreased when increasing the number of macro-layers to $m$=4.
We conjecture that HMFT results with generic $L$$\ge$4 should be recovered for $m$$\ge$6, for which numerical simulations require the use of more sophisticated optimization techniques \cite{Nocedal-Wright_2006,Bergholm_2018,Luo-Yaojl_2020}.

Interestingly, although the variance of the gradient is strongly suppressed with system size, what is commonly considered an indicator of the emergence of barren-plateaux \cite{McClean_2018}, the onset of magnetization (through nonzero mean-field embedding) pushes the convergence of the algorithm within N\'eel and CAF phases when randomly initialized, suggesting an eventual avoidance of barren-plateaux.
On the contrary, the quantum paramagnetic VBS phase (characterized by null mean-fields) is accessed by smoothly tuning the system across the N\'eel-VBS quantum phase transition, as randomly initializing the PQC leads the algorithm to get lost exploring a manifold of states with different orders.
These results highlight the key role of the mean-field embedding on the convergence of the algorithm, and suggest the potential scalability of Q-HMFT to large cluster sizes inaccessible by classical HMFT, i.e. even $L$>4, for which numerical simulation would require highly involved classical algorithms \cite{Markov_2008,Chen_2018,Boixo_2018,DeRaedt_2019}.
Further investigation on the efect of noise on the emergence of barren-plateaux \cite{Wang_2021} would be required to ultimately determine the scalability of the method.

Q-HMFT may be implemented with other technologies, in particular those hosting native XY gates \cite{Kokail_2019} capable to realize different 2D lattices \cite{Monroe_2021}.
Moreover, the simulation of 2D Hamiltonians hosting \textit{exact} VBS ground-states \cite{Schulenburg_2002,Huerga_2017} via Q-HMFT offers a means for benchmarking quantum devices \cite{Kobayashi_2022}.
In addition, the symmetry-guided construction of the PQC makes it suitable for developing error mitigation strategies \cite{Sagastizabal_2019}, or the description of low-energy excitations over the ground-state \cite{Huerga_2013,Higgott_2019}.
From the experimental standpoint, we expect these results to motivate further development and refinement of the family of parameterized XY gates \cite{Salathe_2015,Barends_2019,Foxen_2020} as key elements for variational quantum algorithms.

\begin{acknowledgments} 
We acknowledge useful discussions with Adri\'an Parra-Rodriguez, Alberto Nocera, Polina Feldmann, Christopher Eichler, Gerardo Ortiz, and Robert Raussendorf.
We gratefully acknowledge G. Ortiz for providing access to computing facilities of the Department of Physics, Indiana University.
This research has been funded by the Canada First Research Excellence Fund, Quantum Materials and Future Technologies Program, and the OpenSuperQ project (grant agreement 820363) of the EU Quantum Flagship program.
\end{acknowledgments}

\bibliographystyle{quantum}
\bibliography{vqs_VBS_refs}

\onecolumn

\appendix
\newpage
\section{2$\times$2- and 4$\times$4- cluster-Gutzwiller}
\begin{figure}[b]
\begin{center}
\includegraphics[width=0.6\textwidth]{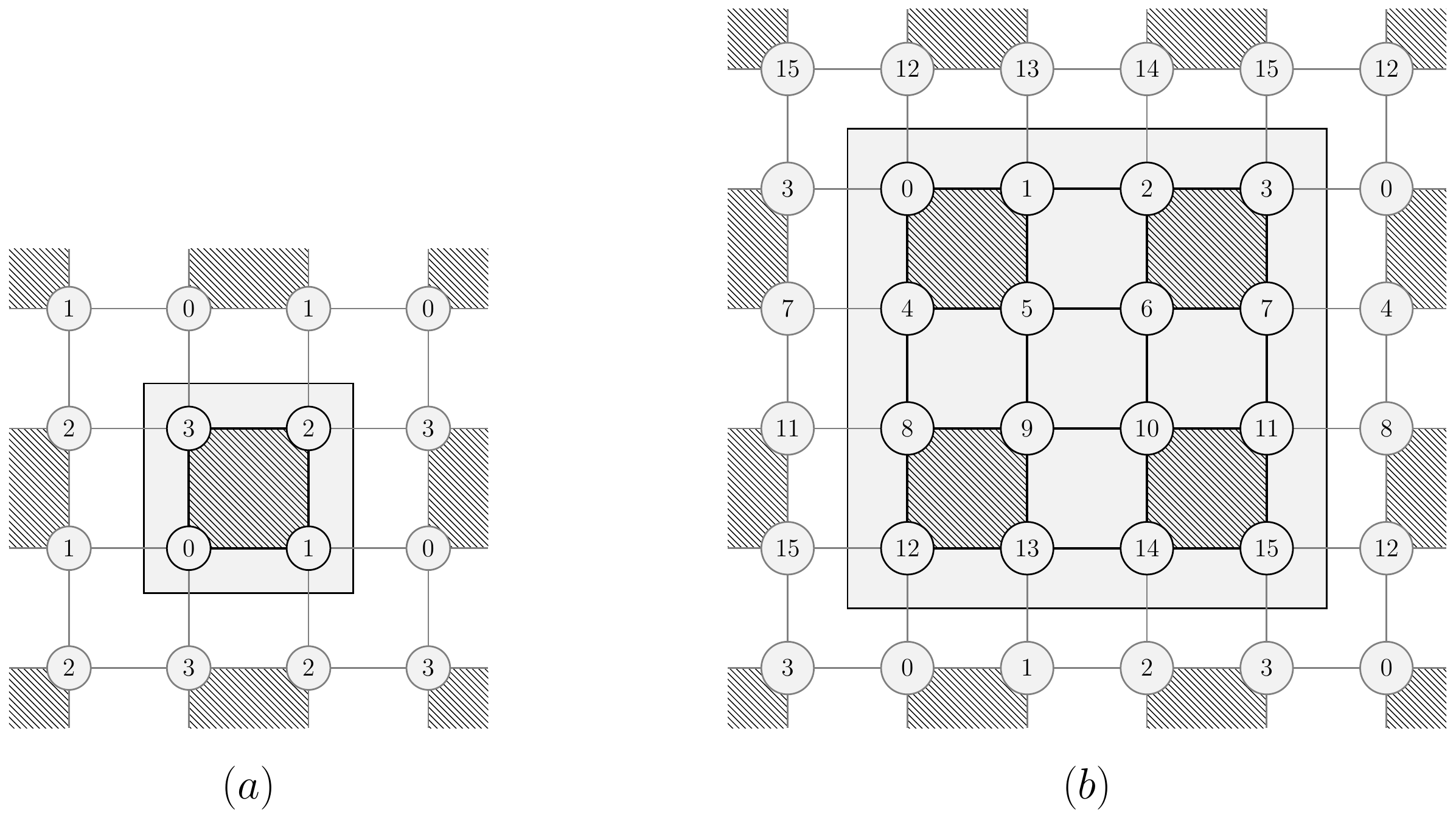}
\end{center}
\caption{\label{fig:clusters}
$(a)$ 2$\times$2 and $(b)$ 4$\times$4 clusters, together with their neighboring sites generating the mean-field embedding.
Hatched squares provide a schematic picture of the \textit{plaquette}-VBS order characterized by the dimer observable.}
\end{figure}

Given a tiling of the lattice and the cluster-Gutzwiller ansatz, the energy per spin of a translational invariant Hamiltoinan reduces to the contribution of the terms acting within the cluster and those connecting different clusters. 
The variational parameters determining the wave function of the cluster are determined through the Rayleigh-Schr\"odinger variational principle.
When considering a \textit{uniform} ansatz (i.e. translational invariant in the coarsed lattice), such energy minimization reduces to minimizing the energy of a single cluster in a self-consistent mean-field embedding (see e.g. \cite{Huerga_2014} of main text).
For the specific case of the 2$\times$2 cluster used in this work (Fig. \ref{fig:clusters}), the energy per spin of the $J_1$--$J_2$ Hamiltonian is
\begin{eqnarray}
E(\bm{\theta})&=& \frac{1}{4} [ J_1\left(
\langle \mathbf{S}_0 \mathbf{S}_1 \rangle 
+ \langle \mathbf{S}_1 \mathbf{S}_2 \rangle 
+\langle \mathbf{S}_2 \mathbf{S}_3 \rangle 
+\langle \mathbf{S}_3 \mathbf{S}_0 \rangle 
\right)
+
J_2\left(
\langle \mathbf{S}_0 \mathbf{S}_2 \rangle 
+ \langle \mathbf{S}_1 \mathbf{S}_3 \rangle 
\right)\notag\\
&&
+J_1 \left(
\langle \mathbf{S}_0\rangle\langle \mathbf{S}_1 \rangle 
+ \langle \mathbf{S}_1\rangle\langle \mathbf{S}_2 \rangle 
+\langle \mathbf{S}_2\rangle\langle \mathbf{S}_3 \rangle 
+\langle \mathbf{S}_3\rangle \langle \mathbf{S}_0 \rangle 
\right)
+
3J_2\left(
\langle \mathbf{S}_0\rangle\langle \mathbf{S}_2 \rangle 
+\langle \mathbf{S}_1\rangle\langle \mathbf{S}_3 \rangle 
\right) ],
\end{eqnarray}
where all expectation values are taken with respect to the wave function of the cluster $\ket{\psi(\bm{\theta})}$. 
Equivalently, the energy per spin obtained with the 4$\times$4 cluster-Guzwiller ansatz,
\begin{eqnarray}
E(\bm{\theta})&=& \frac{1}{16} [ J_1\sum_{\langle i,j\rangle\in\square}
\langle \mathbf{S}_i \mathbf{S}_j \rangle
+ J_2\sum_{\langle\langle i,j\rangle\rangle\in\square}
\langle \mathbf{S}_i \mathbf{S}_j \rangle \nonumber\\
&&+ J_1 (
\langle \mathbf{S}_0 \rangle \langle\mathbf{S}_{12} \rangle 
+\langle \mathbf{S}_1 \rangle \langle\mathbf{S}_{13} \rangle 
+\langle \mathbf{S}_2 \rangle \langle\mathbf{S}_{14} \rangle 
+\langle \mathbf{S}_3 \rangle \langle\mathbf{S}_{15} \rangle 
+ \langle \mathbf{S}_3 \rangle \langle\mathbf{S}_{0} \rangle 
+\langle \mathbf{S}_7 \rangle \langle\mathbf{S}_{4} \rangle \nonumber\\
    &&~~~~+\langle \mathbf{S}_{11} \rangle \langle\mathbf{S}_{8} \rangle 
+\langle \mathbf{S}_{15} \rangle \langle\mathbf{S}_{12} \rangle 
)\nonumber\\
&&+ J_2 (
\langle \mathbf{S}_0 \rangle \langle\mathbf{S}_{13} \rangle 
+\langle \mathbf{S}_1 \rangle \langle\mathbf{S}_{12} \rangle 
+\langle \mathbf{S}_1 \rangle \langle\mathbf{S}_{14} \rangle 
+\langle \mathbf{S}_2 \rangle \langle\mathbf{S}_{13} \rangle
+\langle \mathbf{S}_2 \rangle \langle\mathbf{S}_{15} \rangle 
+\langle \mathbf{S}_3 \rangle \langle\mathbf{S}_{14} \rangle \nonumber\\
&&~~~~
+\langle \mathbf{S}_3 \rangle \langle\mathbf{S}_{12} \rangle 
+\langle \mathbf{S}_0 \rangle \langle\mathbf{S}_{15} \rangle
+\langle \mathbf{S}_3 \rangle \langle\mathbf{S}_{4} \rangle 
+\langle \mathbf{S}_7 \rangle \langle\mathbf{S}_{0} \rangle 
+\langle \mathbf{S}_7 \rangle \langle\mathbf{S}_{8} \rangle 
+\langle \mathbf{S}_{11} \rangle \langle\mathbf{S}_{4} \rangle\nonumber\\
    &&~~~~
+\langle \mathbf{S}_{11} \rangle \langle\mathbf{S}_{12} \rangle
+\langle \mathbf{S}_{15} \rangle \langle\mathbf{S}_{8} \rangle
)].
\end{eqnarray}

If the parameterization of the wave function is lineal, i.e. $\ket{\psi{(\bm{\theta})}}=\sum \theta_{ \lbrace S^z_j\rbrace} \ket{\lbrace S^z_j\rbrace}$ with $\theta\in\mathbb{C}$ uncovers the whole Hilbert space of the cluster. 
Equating the derivative of the energy to zero, 
$
\partial_{\theta_k} E(\bm{\theta})=0, 
$
leads to a non-lineal set of equations that can be cast in matrix form, and solved by iteratively performing exact diagonalization of the $L$$\times$$L$ cluster with open boundary conditions and a set of self-consistent mean-fields, $\bra{\psi(\bm{\theta})} \mathbf{S}_j\ket{\psi(\bm{\theta})}$, acting on its boundaries.
LRO is generically identified by a non-zero mean-field embedding, i.e. $\langle \mathbf{S}_j\rangle$$\ne$0, while quantum paramagnetic phases are generically identified by optimal solutions where the mean-field embeddings is null, i.e. $\langle \mathbf{S}_j \rangle$=0.

Restriction to the null magnetization subspace, $\sum_{j\in\square} \langle S^z_j\rangle=0$, restricts the onset of magnetizations along $z$, i.e. $\langle S^{x,y}_j\rangle=0$, and allows to increase the efficiency of the computation while still allowing for the eventual breakdown of SU(2) and onset of long-range order (LRO).
Nevertheless, exact diagonalization is limited by classical computing ressources to $N\lesssim 30$.

\section{U(1) Parameterized Quantum Circuits}
\begin{figure}[t]
\begin{center}
\includegraphics[width=1\textwidth]{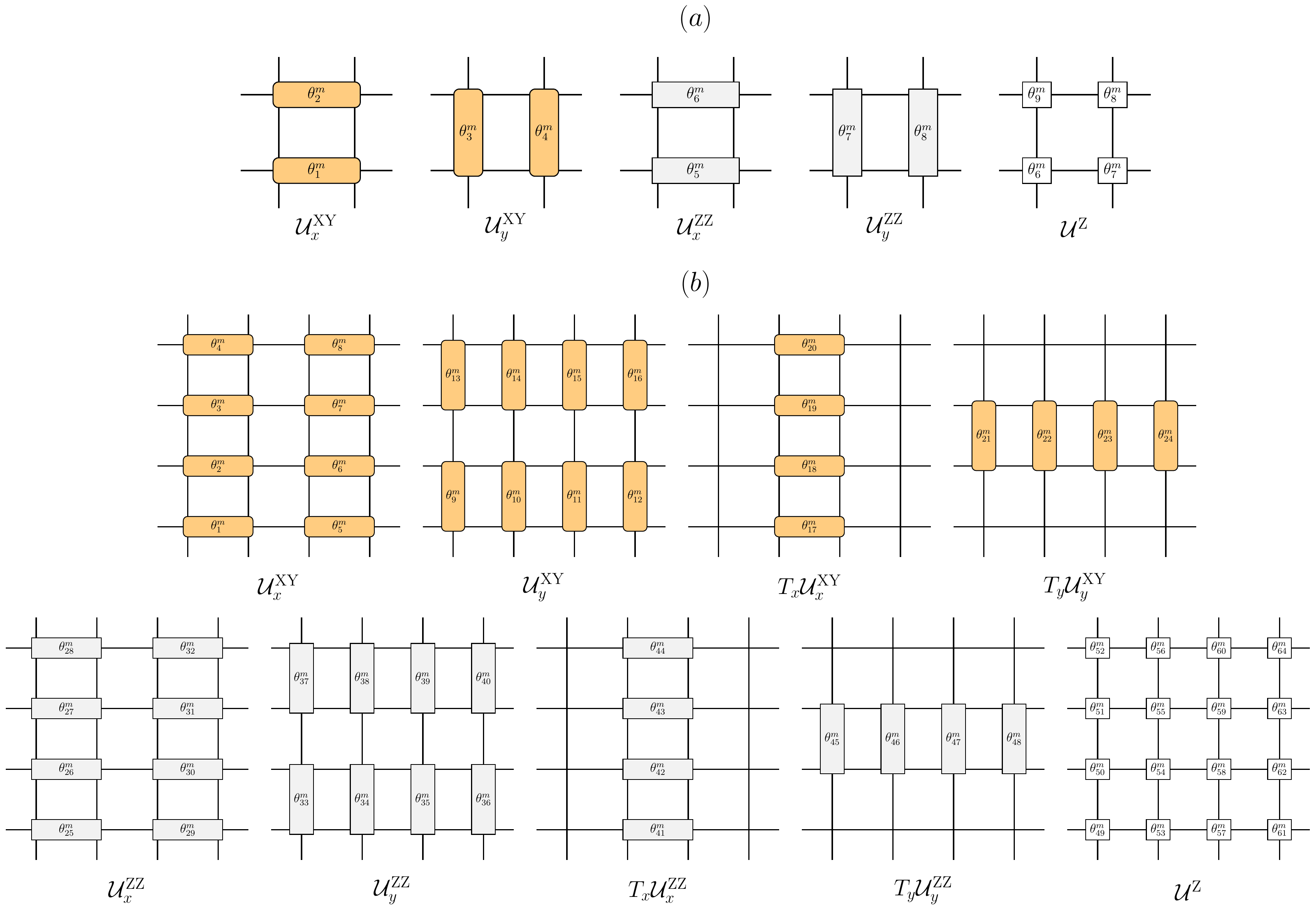}
\end{center}
\caption{
\label{fig:PQC}
Ordered set of layers of {\sf REAL}-XY gates (orange rounded rectangles), ZZ- (gray rectangles), and Z-rotations (white squares) comprising a \textit{macro-layer} of the PQC for ($a$) 2$\times$2 and ($b$) 4$\times$4 clusters, where $\theta_k^m$ labels the $k$th variational parameter within the $m$th macro-layer.
Generalization to even $L$$\times$$L>4$$\times$$4$ is straightforward and has the same depth as $L$=4, $d$=$9m$.
}
\end{figure}

As described in the main text, the PQC is constructed by a recursive application of two-qubit XY gates respecting nearest-neighbor (NN) connectivities in an order that, given that gateset, minimizes the circuit depth and favors the C$_4$ symmetry of the square lattice as much as possible (see Fig. \ref{fig:PQC}).
Additional layers of two-qubit ZZ-rotations, applied in the same order described before, and a final layer of single-qubit Z-rotation gates allow to tune the sign-structure of the wave function and add correlations.
Such a XY-ZZ-Z macro-layer structure is repeated $m$ times with independent parameters to increase numerical accuracy.

As few as $m$=2 macro-layers are sufficent to provide with a good approximation to the classical cluster-Gutzwiller results, where the cluster wave function uncovers the whole Hilbert space of the cluster.
In particular, for 2$\times$2 the hybrid Q-HMFT numerical results reproduce exactly the classical cluster-Gutziller results with $m$=2 and restricting to a single parameter within the XY and ZZ layers, respectively, i.e. $\theta_1^m$=$\theta_2^m$=$\theta_3^m$=$\theta_4^m$ and $\theta_5^m$=$\theta_6^m$=$\theta_7^m$=$\theta_8^m$, where we have defined $\theta^m_k$ as the $k$th parameter within the macro-layer $m$.

\section{Convergence}
Although the variance of the gradient of the energy is strongly suppressed well within the three main phases identified in the $J_1$--$J_2$ Heisenberg AFM, i.e. at $J_2$=0, 0.5, 1, the noiseless numerical simulations of Q-HMFT with $m$=2 show that the convergence of the Q-HMFT algorithm is pushed by the onset of long-range order (see Fig. \ref{fig:convergence} in main text), eventually escaping from noise-free barren-plateaux.
On the contrary, within quantum paramagnetic phases, the system does not arrive to converge and explores states closely in energy but with different orders.

\begin{figure}[t]
\begin{center}
\includegraphics[width=0.6\textwidth]{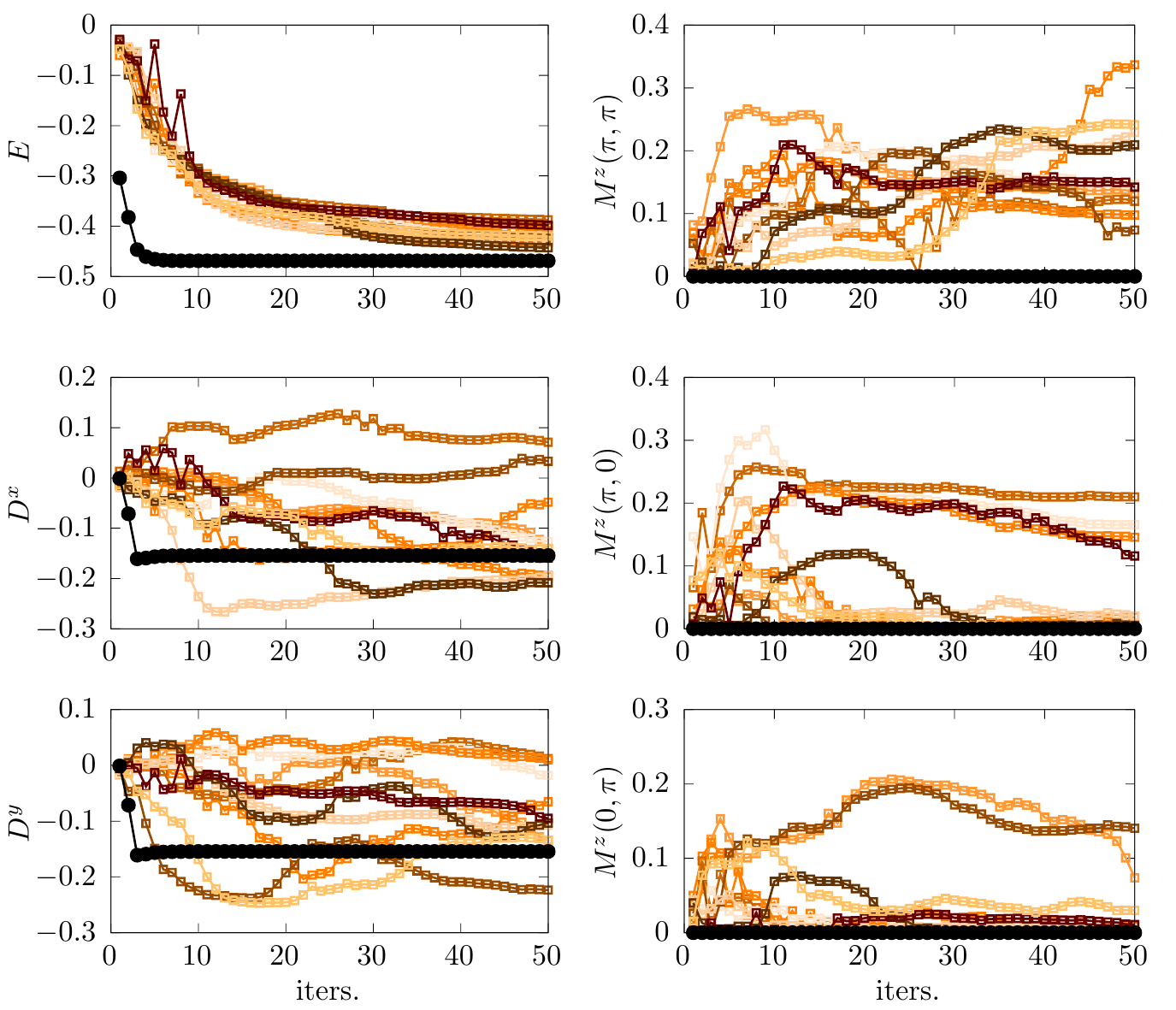}
\end{center}
\caption{\label{fig:convergence_VBS}
Energy, magnetization, and dimer observables of Q-HMFT-4$\times$4, $m$=2, versus number of iterations well within the quantum paramagnetic phase, i.e. at $J_2$=0.5, for 10 random initializations of the PQC (empty squares, orange scale) together with classical HMFT results from a random initialization (black dots) for comparison.
}
\end{figure}

Within the HMFT framework, quantum paramagnetic phases are generically identified by converged solutions with null mean-field embeddings (see e.g. Refs. \cite{Isaev_2009_J1J2,Huerga_2014,Huerga_2016} of main text).
In general, in the classical cluster-Gutzwiller algorithm, the system is able to find optimal quantum paramagnetic solutions, even if the initial seed breaks the symmetries (in this case SU(2)).
However, in Q-HMFT the system explores a manifold of states close in energy but with different orders, not being able to stabilize a quantum paramagnetic state with well-defined dimerization (see Fig. \ref{fig:convergence_VBS}).

\end{document}